# Partial compensation of thermal noise in the fundamental mode of an optical cavity

N. O. Zhadnov[1], N. N. Kolachevsky[1,2]

[1] P.N. Lebedev Physical Institute of the Russian Academy of Sciences, Leninskiy Prospekt 53, 119991, Moscow

[2] Russian Quantum Center, Skolkovo Innovation Center, Bolshoi Boulevard, 30, bldg. 1, 121205, Moscow

e-mail: nik.zhadnov@yandex.ru

*Thermal noise of optical cavities limits the accuracy of many experiments on precision laser spectroscopy and interferometry. The study of the physical properties of this noise opens opportunities for creation of more stable cavities, reduction of phase noise of optical radiation and performance of accurate optical studies. The paper proposes a method of partial recording of $TEM_{00}$ mode thermal noise of a Fabry-Perot cavity using two "probe" modes of higher order. The method allows partial compensation of noise in the fundamental mode. Mathematical modeling is performed, which confirms the efficiency of the method.*

## 1. Introduction

Highly-stable optical cavities are widely used in experiments related to precision optical measurements: for registration of gravitational waves [1], searches for dark matter [2,3], high-resolution spectroscopy in optical atomic clocks [4,5], and in other fundamental research [6,7] Stabilization of laser radiation frequency by reference monolithic high-Q Fabry-Perot cavities using the Pound-Driver-Hall (PDH) technique [8] allows to reduce spectral linewidth to values less than 100 mHz [9–11].

The instability of the cavity mode frequency ν is related to the instability of its length L by the relation:

$$\frac{\Delta \nu}{\nu} = -\frac{\Delta L}{L}. \qquad (1)$$

The fundamental stability limit of L is determined by the Brownian thermal fluctuations of the cavity, which are commonly called thermal noise. Their physical properties can be described by the fluctuation-dissipation theorem [12,13]. Random oscillations of this nature are inherent to any heated body. Thermal noise of a Fabry-Perot cavity can be represented as random vibrations of the mirror surface, averaged over the light mode spot on each mirror, leading to fluctuations of cavity length. To date, the maximum achievable frequency stability of the best laser systems with reference cavities is limited by thermal noise. The sensitivity of LIGO gravitational wave detectors in the frequency range of 30–200 Hz is also limited by thermal fluctuations of the interferometer mirrors [14]. An efficient method for calculating mirror thermal noise was proposed in [15].

Significant progress in reducing thermal noise can be achieved by increasing the length of the cavity, lowering its operating temperature, and using materials with a high mechanical quality factor (quartz, single-crystal silicon) for the manufacture of mirrors [16]. It is also possible to reduce thermal noise limit by approximately a factor of two by using other higher-order Gaussian modes instead of the fundamental TEM$_{00}$ mode [17]. This effect is achieved due to the large transverse dimensions of higher-order modes, which leads to better averaging of thermal surface fluctuations. In the presence of several ultrastable laser systems with close frequency instability characteristics, it is possible to average their frequencies and thus overcome the thermal noise limit for a single system [18].

In this paper, we propose a method to measure the thermal noise of a selected "main" mode of the cavity by excitation of two additional "probe" modes in it, which makes it possible to obtain an RF signal whose frequency noise contains information about the thermal noise of the "main" mode. Then, this signal can be used to partially compensate laser frequency noise (in the main mode), by analogy with the vibration compensation method described in [19]. The presence of a thermal frequency response of cavity to a change in the radiation power fed into it [9,20] opens opportunities for active suppression of thermal noise. Furthermore, the measurement and analysis of thermal fluctuations will make it possible to study the mechanical properties of multilayer Bragg mirrors [21].

## 2. Partial measurement of thermal noise of a cavity mode frequency

Since the laser mode has finite dimensions on the cavity mirrors, only those parts of the reflecting surface on which the radiation is incident contribute to the thermal noise. It can be shown that the phase shift of a beam with an intensity profile normalized to unity $g(\vec{r})$ reflected from the mirror surface perturbed by thermal noise $\vec{u}(\vec{r},t)$ (Fig. 1) is expressed as [22]:

$$\Delta\phi(t) = \int_S g(\vec{r}) \cdot (\vec{k}, \vec{u}(\vec{r},t)) \cdot d^2r. \qquad (2)$$

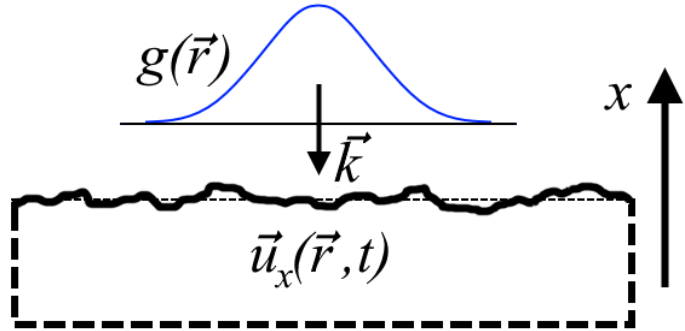

*Figure 1. Gaussian beam of the TEM$_{00}$ mode incident on a mirror surface distorted by thermal noise.*

Integration, in this case, is performed over the surface of the mirror S; $\vec{r}$ is the radius vector of a point on the mirror surface; $\vec{u}(\vec{r},t)$ is the displacement of the mirror point with the radius vector $\vec{r}$ from the unperturbed position; $\vec{k}$ is the wave vector of the incident radiation perpendicular to the mirror surface. Using the phase shift $\Delta\phi(t)$, we can introduce the value of the "effective" displacement of the mirror surface along the $x$ axis:

$$X(t) = \frac{\Delta\phi(t)}{|\vec{k}|} = \int_S g(\vec{r}) \cdot u_x(\vec{r},t) \cdot d^2r. \qquad (3)$$

Any cavity eigenmode can be represented as a superposition of Hermite-Gaussian or Laguerre-Gaussian functions [23]. Due to the more efficient averaging of the fluctuations of the reflecting surface, modes whose intensity is distributed over a larger area of the mirror are less sensitive to thermal noise. The intensity distribution $g(\vec{r})$ for each mode has a specific shape, because of which thermal noise for different modes can be determined by thermal fluctuations in non-intersecting parts of the mirror. The proposed method of thermal noise measurement is based on this fact.

Its implementation requires to select three modes: M1 is the "main" one, that is, the one whose thermal noise needs to be partially characterized, and M2 and M3 are "probe". We assume that all three laser beams corresponding to the modes are frequency-locked to the cavity and the frequency stability of each of them is limited by thermal noise. To avoid interference effects, all modes must have different frequencies, which can be implemented using a single laser by performing the required frequency detuning using modulators. Let us divide the mirror surface into two conditional regions: O1, where the intensity of the main mode M1 is concentrated, and O2, where the intensity of the probe mode M2 is concentrated. In this case, the M2 mode should be chosen in such a way that it overlaps with the M1 mode as little as possible. In turn, the M3 mode should have a strong overlap with the M1 and M2 modes. An example of three modes that meet these requirements is the Laguerre-Gaussian modes $LG_{00}$ (M1), $LG_{03*}$ (M2), $LG_{10}$ (M3), shown in Fig. 2 (see also supplementary materials). $LG_{03*}$ belongs to the Laguerre-Gaussian modes with an orbital angular momentum [17].

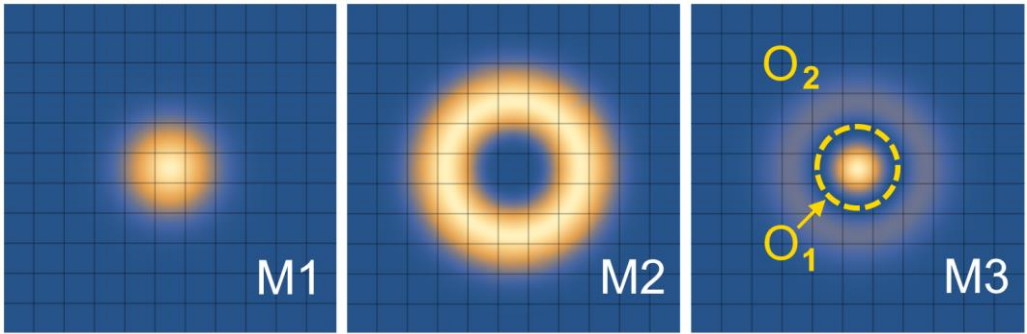

*Figure 2. Laguerre–Gaussian mode profiles. From left to right: the fundamental mode $LG_{00}$ (M1), and probe modes $LG_{03*}$ (M2), $LG_{10}$ (M3). The regions O1 and O2 described in the text are also indicated. Intensity profiles of these modes are shown on Fig. 4a.*

When the optical fields of the M2 and M3 modes are heterodyned, their frequency thermal fluctuations (that arise from surface fluctuations in the O2 region) are correlated and will be partially compensated in the beat signal (radio frequency range). In this case, the beat signal will mostly contain thermal noise created by vibrations of the mirror surface in the region O1, which is covered by the fundamental mode M1 ($LG_{00}$). Thus, frequency fluctuations of the received signal can be subtracted from the frequency of the M1 mode, for example, using an acousto-optic modulator (AOM) (Fig. 3). This reduces the contribution of thermal noise to the frequency instability of laser radiation in the fundamental mode M1. In this case, the residual frequency fluctuations $\delta\vartheta_{132}$ will be:

$$\delta\vartheta_{132}(t) = -\frac{\vartheta_1}{L}\int_S [g_1(\vec{r}) - (g_3(\vec{r}) - g_2(\vec{r}))] \cdot u_x(\vec{r},t) \cdot d^2r. \qquad (4)$$

## 3. Modelling

The method described above was analyzed using mathematical modeling. To do this, the surface $u_x(\vec{r})$ corresponding to Brownian thermal vibrations was repeatedly randomly created, and then, in accordance with formula (3), the integral displacements of the surface were calculated for the intensity profile of each mode. For simplicity, only the thermal noise of one reflecting surface in the static case is considered in the calculation. The role of a mirror perturbed by thermal noise was played by a fractal surface formed as a result of a Brownian process. The method for creating such surfaces is described in [24]. The resulting functions $u_x(\vec{r})$ are described using two parameters: the lacunarity $r$ and the Hurst coefficient $h$. The last parameter is mostly responsible for the "noisiness" of the surface. For modeling of thermal noise, which is a $1/f$ noise, $h$ values close to 1 are most suitable [25]. The lacunarity parameter was taken equal to 0.5 for all calculations.

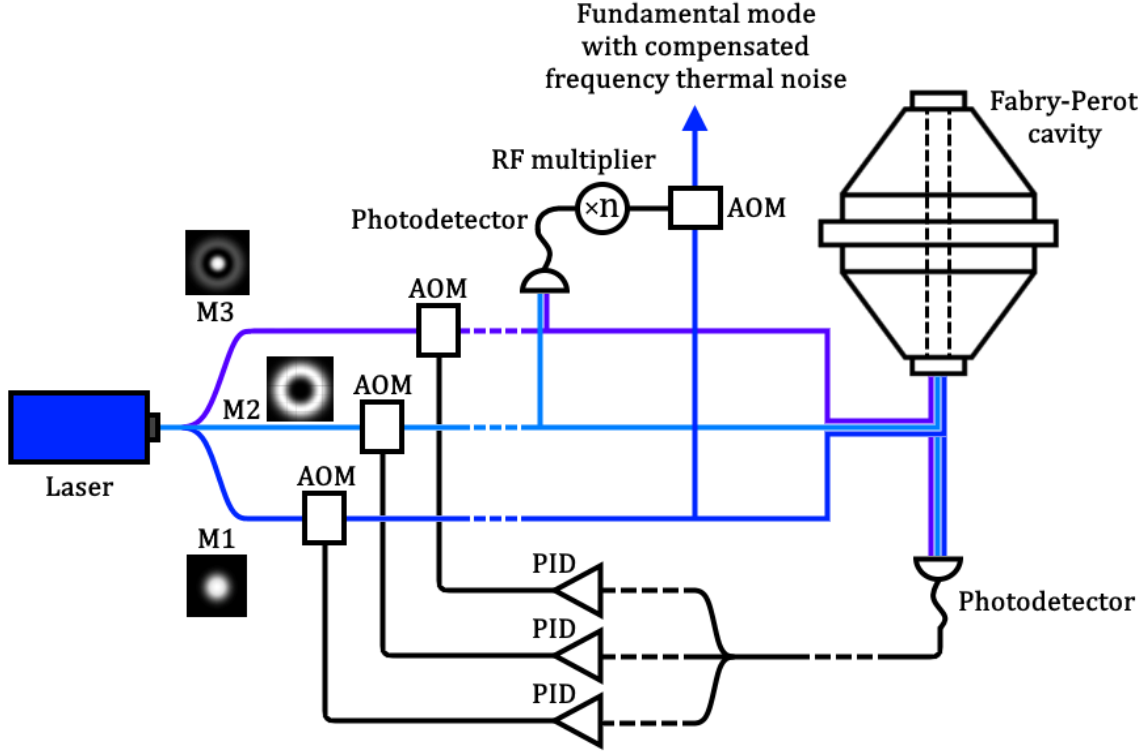

*Figure 3. Simplified experimental scheme for measurements and partial compensation of the frequency thermal noise of the laser light in the fundamental mode M1. Frequencies of all laser beams are locked to corresponding modes M1, M2, M3. Some details of PDH schemes are not shown. Three AOMs are used for frequency locking. The beat signal between M2 and M3 is used for the frequency thermal noise compensation of M1 with the help of another AOM.*

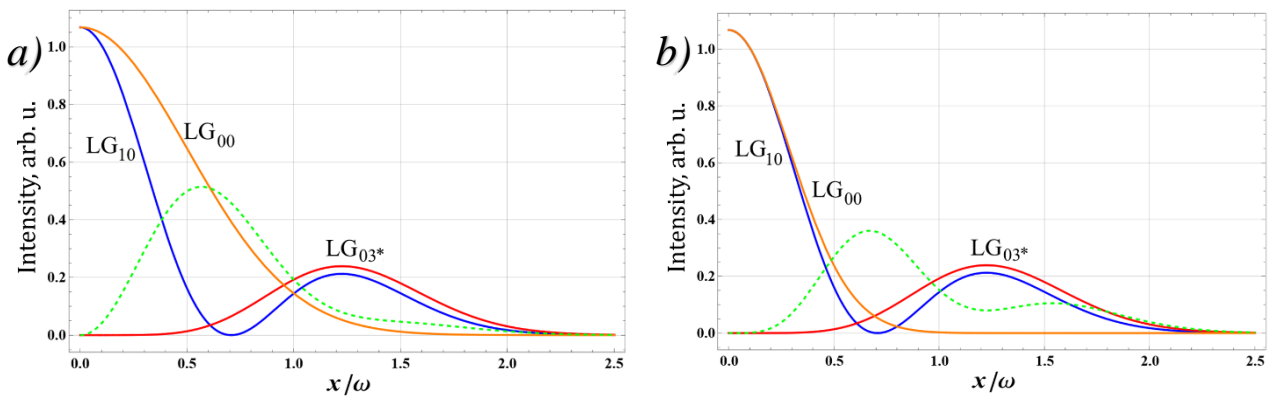

*Figure 4. (a) Intensity profiles of modes shown in Fig. 2. The orange curve (fundamental mode M1, $LG_{00}$) corresponds to the function $g_1(x/\omega)$, the red curve (probe mode M2, $LG_{03*}$) corresponds to $g_2(x/\omega)$, and the blue curve (probe mode M3, $LG_{10}$) corresponds to $g_3(x/\omega)$. The green dashed curve corresponds to the combination $[g_1(\vec{r}) - (g_3(\vec{r}) - g_2(\vec{r}))]$. $\omega$ is the radius of the spot of the fundamental mode $LG_{00}$ on the mirror by the intensity level $1/e^2$. (b) Plot like (a), but the mode $LG_{00}$ has a three times shorter wavelength than $LG_{10}$, $LG_{03*}$. The mode $LG_{00}$ intensity in the figure (orange*

curve) is decreased by a factor of 3 to show the coincidence with the central spot profile of the mode LG$_{10}$ clearly. The green dashed curve corresponds to the combination $[g_1(\vec{r}) - 3*(g_3(\vec{r}) - g_2(\vec{r}))]$.

In the calculations, Brownian surfaces of various noisiness were used, with Hurst coefficient values in the range of 0.5-1 (Table 1). For each value, 15 random surfaces were generated and $X_i$ calculations were performed using formula (3) for each of the three modes $i = 1,2,3$ (Figure 5). The obtained values of $X_i$ made up 15-term samples, for which the standard deviations $\sigma_{Xi}$ were calculated. Root-mean-square deviations of the surface $\sigma_{X132}$ for combinations $X_{132} = X_1 - (X_3 - X_2)$ were also calculated, the values of which allow to judge the effectiveness of the proposed method of thermal noise compensation. The data obtained are presented in Table 1.

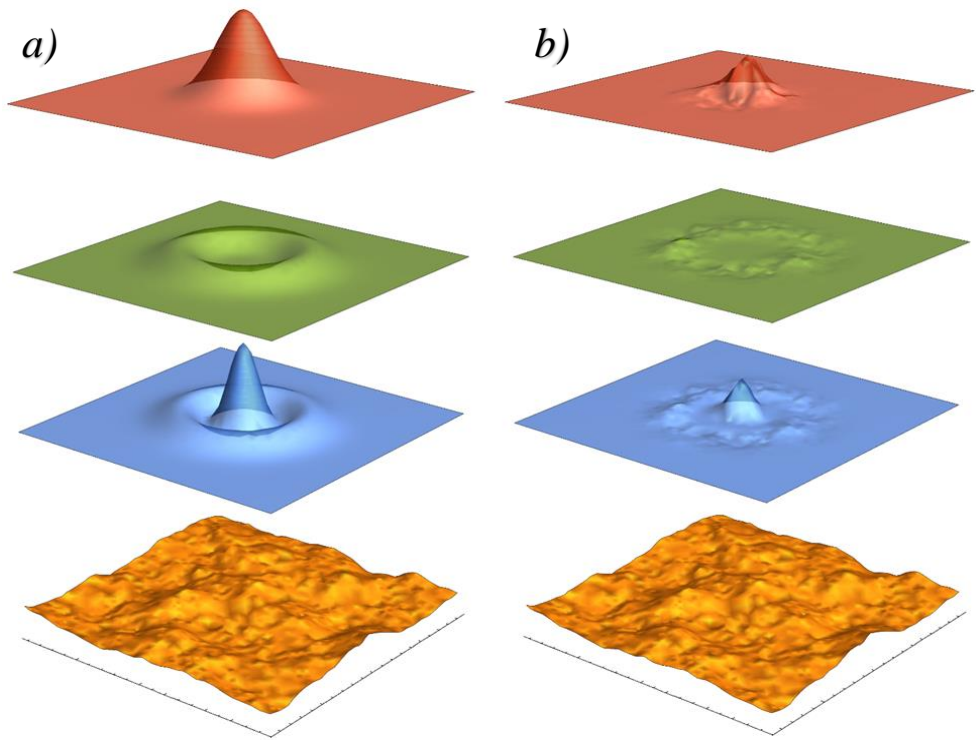

*Figure 5. Illustration of thermal noise for three modes. (a) Spatial intensity profiles of modes (top—down) LG$_{00}$ (M1), LG$_{03*}$ (M2), LG$_{10}$ (M3) and Brownian surface $u_x(\vec{r})$ with $h = 0.8$. (b) Functions obtained by multiplying corresponding intensity profiles by $u_x(\vec{r})$.*

| h | $\sigma_{X1}$ | $\sigma_{X2}$ | $\sigma_{X3}$ | $\sigma_{X132}$ | $\sigma_{X132}/\sigma_{X1}$ |
|---|---|---|---|---|---|
| 0.5 | 11.47 | 7.52 | 9.19 | 9.35 | 0.82 |
| 0.6 | 11.29 | 6.13 | 6.95 | 8.40 | 0.75 |
| 0.7 | 9.27 | 4.46 | 5.66 | 6.74 | 0.73 |
| 0.8 | 4.01 | 2.56 | 2.68 | 3.14 | 0.79 |
| 0.9 | 3.15 | 1.58 | 1.27 | 1.72 | 0.55 |
| 0.94 | 3.68 | 1.40 | 2.25 | 2.54 | 0.69 |
| 0.95 | 2.63 | 0.86 | 1.33 | 1.79 | 0.68 |
| 0.98 | 1.52 | 0.70 | 0.90 | 1.12 | 0.74 |

*Table 1. Calculation results for the case where all modes have the same wavelength.*

Since the characteristic values of the humps and dips of the Brownian surfaces used in the calculation are not related to the physical properties of real mirrors, it is not the absolute values of the calculated values that are indicative, but their ratios. As can be seen from the last column of Table 1, the ratio $\sigma_{X132}/\sigma_{X1}$ fluctuates around 0.7, which indicates that as a result of subtracting the noise difference of the probe modes from the fundamental mode, its noise decreases. This decrease turns out to be relatively small due to insufficiently good matching of the mode profiles (Fig. 4a).

It is possible to improve the matching of the selected mode profiles by selecting the radiation wavelength $\lambda$ since the radius of the Gaussian beam ω is proportional to $\sqrt{\lambda}$. By reducing the wavelength of the M1 mode by a factor of 3, it is possible to achieve a fairly accurate coincidence of its contour and the contour of the central spot of the M3 mode, as shown in Fig. 4b. In this case, the beat signal of the probe modes more accurately describes the thermal noise of the central region of the mirror, in which the intensity of the M1 mode is concentrated. The results of the corresponding calculations are presented in Table 2.

| h | $\sigma_{X1}$ | $\sigma_{X2}$ | $\sigma_{X3}$ | $\sigma^*_{X132}$ | $\sigma^*_{X132}/\sigma_{X1}$ |
|---|---|---|---|---|---|
| 0.5 | 14.49 | 6.62 | 6.86 | 6.66 | 0.46 |
| 0.6 | 12.21 | 4.95 | 4.73 | 4.41 | 0.36 |
| 0.7 | 11.41 | 4.35 | 5.90 | 5.39 | 0.47 |
| 0.8 | 6.68 | 2.89 | 3.22 | 3.32 | 0.50 |
| 0.9 | 5.59 | 1.04 | 2.01 | 1.64 | 0.29 |
| 0.94 | 2.99 | 1.04 | 1.36 | 1.19 | 0.40 |
| 0.96 | 2.54 | 0.98 | 1.26 | 1.16 | 0.46 |
| 0.98 | 1.49 | 1.11 | 0.94 | 0.99 | 0.66 |

*Table 2. Calculation results for the case where modes M1 and M2 have the same wavelength, and the wavelength of the mode M3 is three times shorter.*

The calculations of $\sigma^*_{X132}$ were made for the quantity $X^*_{312} = X_1 - 3*(X_3 - X_2)$, which considers the ratio of the intensities of the M3 and M1 modes at the center of the mirror, equal to 3. In the experiment, a threefold increase in the frequency of the beat signal M3 and M2 can be provided with a frequency multiplier (Fig. 3). The obtained values of $\sigma^*_{X132}/\sigma_{X1}$ in the vicinity of 0.5, indicate that the improvement in geometric matching leads to more efficient noise compensation.

Similar calculations were carried out with a different kind of random surfaces: with B-splines built on a matrix of points with a random coordinate. The results obtained indicate that the method works with approximately the same efficiency as in the case of Brownian surfaces.

## 4. Conclusions

The proposed method of partial compensation of frequency thermal fluctuations for a laser, locked to the fundamental mode of a reference cavity, is based on the fact that the thermal noise of each individual mode is determined by the noise of that region of the mirror surface, on which its intensity is concentrated. Mathematical modeling of Brownian surfaces made it possible to calculate the average displacements of mirrors due to the Brownian surface motion for different modes and confirm that the beat signal of the $LG_{10}$ and $LG_{03*}$ modes partially contains the thermal noise of the $LG_{00}$ mode. The choice of the main $LG_{00}$ mode is not accidental: most stabilized lasers and gravitational detectors operate on this mode, since this provides the best coupling with the laser mode and, accordingly, the maximum power. In this case, higher-order probe modes, which have a lower level of thermal noise, are auxiliary and provide the ability to compensate for fluctuations in the main mode.

In the case when all modes have the same wavelength, the method allows compensation of about 30% of the thermal noise in the fundamental mode. If the fundamental mode has a triple frequency, then the coincidence of the intensity profiles is better, and the subtraction allows to suppress of about 50% of the noise. In this case, the characteristic values of the standard deviation $\sigma_{X132}$ of the mirror surface for the fundamental mode, even after subtracting some of the thermal noise, turn out to be a bit larger than the values of $\sigma_X$ corresponding to the probe modes $LG_{10}$ and $LG_{03*}$. Higher-order modes are difficult to efficiently excite, so the method can be used to increase the frequency stability of the radiation corresponding to the selected operating mode. The three modes analyzed were chosen to illustrate the method and are not likely to be the optimal combination. A more optimal choice of modes with better matching of intensity profiles will make it possible to compensate even better for noise in the fundamental mode.

FUNDING. This work was supported by the Russian Foundation for Basic Research (grant no. 19-32-90207).

## SUPPLIMENTARY

| LG$_{00}$ | $g_1(r,\varphi) = \tilde{g}_1 \cdot \exp\left(-\dfrac{2r^2}{w^2}\right)$ |
|---|---|
| LG$_{03*}$ | $g_2(r,\varphi) = \tilde{g}_2 \cdot \left(1 - \dfrac{2r^2}{w^2}\right)^2 \cdot \exp\left(-\dfrac{2r^2}{w^2}\right)$ |
| LG$_{10}$ | $g_3(r,\varphi) = \tilde{g}_3 \cdot \left(\dfrac{r}{w}\right)^6 \cdot \exp\left(-\dfrac{2r^2}{w^2}\right)$ |

*Intensity functions for the three modes used in the calculation. The coefficients $\tilde{g}$ are responsible for the normalization to 1.*